\documentclass[conference]{IEEEtran}
\usepackage{amsmath}
\usepackage{latexsym}
\usepackage{graphicx}
\usepackage{bbding}
\usepackage{indentfirst}
\usepackage{algorithm,algorithmic}
\usepackage{setspace}
\usepackage{float}
\usepackage{epstopdf}
\usepackage{amssymb}
\usepackage{amsfonts}
\usepackage{enumerate}
\usepackage{multicol}
\usepackage{color}
\usepackage{slashbox}
\usepackage{bm}
\usepackage{amssymb}
\usepackage{stfloats}
\IEEEoverridecommandlockouts
\allowdisplaybreaks[4]

\begin{document}
\title{Covariance-Based Cooperative Activity Detection for Massive Grant-Free Random Access}
\author{\IEEEauthorblockN{Xiaodan Shao$\IEEEauthorrefmark{1}$, Xiaoming Chen$\IEEEauthorrefmark{1}$, Derrick Wing Kwan Ng$\IEEEauthorrefmark{2}$, Caijun Zhong$\IEEEauthorrefmark{1}$, and Zhaoyang Zhang$\IEEEauthorrefmark{1}$}

\IEEEauthorblockA{$\IEEEauthorrefmark{1}$College of Information Science and Electronic Engineering, Zhejiang University, Hangzhou, China\\
$\IEEEauthorrefmark{2}$ School of Electrical Engineering and Telecommunications, University of New South Wales, Sydney, Australia\\
E-mails: \{shaoxiaodan, chen\_xiaoming, caijunzhong, ning\_ming\}@zju.edu.cn, w.k.ng@unsw.edu.au
}}\maketitle

\begin{abstract}
This paper designs a cooperative activity detection framework for massive grant-free random access in the sixth-generation (6G) cell-free wireless networks based on the covariance of the received signals at the access points (APs). In particular, multiple APs cooperatively detect the device activity by only exchanging the low-dimensional intermediate local information with their neighbors. The cooperative activity detection problem is non-smooth and the unknown variables are coupled with each other for which  conventional approaches are inapplicable. Therefore, this paper proposes a covariance-based algorithm by exploiting the sparsity-promoting and similarity-promoting terms of the device state vectors among neighboring APs. An approximate splitting approach is proposed based on
the proximal gradient method for solving the formulated problem. Simulation results show that the proposed algorithm is efficient for large-scale activity detection problems while requires shorter pilot sequences compared with the state-of-art algorithms in achieving the same system performance.
\end{abstract}

\begin{IEEEkeywords}
Cooperative activity detection, massive access, 6G cell-free wireless networks, covariance-based detection.
\end{IEEEkeywords}

\IEEEpeerreviewmaketitle
\section{Introduction}
The massive machine-type communications (mMTC), which is a typical application scenario for 6G wireless networks, aims to meet the demand for massive connectivity for hundreds of billions of Internet-of-Things (IoT) devices. For the massive access, conventional grant-based random access schemes lead to an exceedingly long access latency and a prohibitive signaling overhead. To this end, grant-free random access schemes have been considered as a promising candidate technique for realizing 6G cellular IoT \cite{ma}, where active devices transmit their data signals without obtaining a grant from the base station (BS) after sending pre-assigned pilot sequences. Hence, the key to grant-free random access is active device detection at the BS based on the received pilot sequences \cite{5gderrick}.

Inspired by the sporadic characteristics of IoT applications, several compressed sensing-based approaches have been proposed to detect active devices for grant-free random access systems. For instance, in \cite{amp} and \cite{amp1}, the approximate message propagation (AMP) algorithms were designed for activity detection in different scenarios by exploiting the statistics of wireless channels. However, the AMP algorithms require high computational complexity. As a result, the authors in \cite{shaodim} proposed a low-complexity dimension reduction-based algorithm, which projects the original device state matrix to a low-dimensional space by exploiting its sparse and low-rank structure. Note that the above approaches in \cite{amp}-\cite{shaodim} perform activity detection based on the instantaneous received signals. Recently, a covariance-based algorithm has been proposed to improve the performance of device activity detection in \cite{covar}, where the detection problem was solved by a coordinate descent algorithm with random sampling. In general, these algorithms, e.g. \cite{amp}-\cite{covar}, exploiting the sparsity structure of the device state matrix, which enjoy reasonable detection performance. However, due to a large number of devices and the limited radio resources in $6$G networks for massive access, the active device detection has been emerging as a challenging problem.

To overcome this challenge, multi-cell massive access with multiple APs were applied to the problem of active device detection. For example, the multi-cell sparse activity detection was proposed in \cite{multicell}, where each AP operates independently to perform activity detection and channel estimation for the devices distributed in its own cell by treating the inter-cell interference as noise. In fact, if the APs can jointly process the pilot sequences received from the devices in neighboring APs, the detection performance can be further improved even with only short pilot sequences. Motivated by this fact, this paper considers a 6G cell-free wireless network, where multiple APs deployed in a vast area to serve all devices located in this area \cite{gaozhen1, antennaderrick}. In particular, cooperative activity detection among the APs requires extra information exchanges in the system. To reduce the amount of associated signaling overhead, this paper designs a scalable computationally efficient algorithm to detect the active devices, which is reliable and robust to AP and/or backhaul link failure and the variation in channel statistics. The main contributions of this paper are as follows:
\begin{enumerate}

\item The paper proposes a novel cooperative activity detection framework for grant-free random access in 6G cell-free wireless networks based on the covariance of the received signals.

\item This paper proposes a cooperative massive detection (CMD) algorithm by exploiting the special characteristic of the device state vectors of interest among the neighboring APs, namely joint similarity and sparsity.

\item This paper analyzes the computational complexity and the communication cost of the proposed CMD algorithm which shows its effectiveness in 6G cell-free wireless networks.

\end{enumerate}

\section{System Model}
\begin{figure}[t]
  \centering
\includegraphics [width=0.333\textwidth] {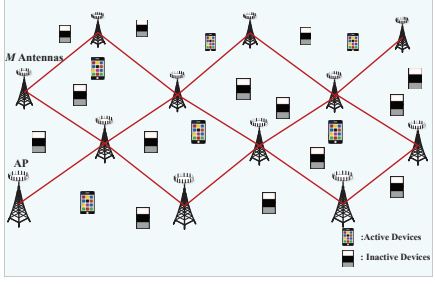}
\caption{Illustration of a 6G cell-free wireless network with multiple APs.}
\label{cooperative}
\end{figure}

Consider a 6G cell-free wireless network comprising $B$ APs. The APs are equipped with $M$ antennas each, serving $N$ uniformly distributed single-antenna IoT devices in a vast area. Each AP is connected to several adjacent APs via backhaul links and can only communicate with its one-hop neighbors for reducing the communication load, as shown in Fig. \ref{cooperative}. Due to the burst characteristic of IoT applications, only a fraction of IoT devices are active at any given time slot. Let $|\cdot|_c$ denote the cardinality of a set. We use $\mathcal{K}$ to denote the set of active devices with $K=\left|\mathcal{K}\right|_c\ll N$ being the number of active devices. For convenience, we define $\chi_n$ as the binary activity indicator with ${\chi_n} = 1$ if the $n$th device is active, and ${\chi_n} = 0$ otherwise. Moreover, we represent the $M$-dimensional channel vector from the $n$th device to the $b$th AP as $\sqrt{g_{b,n}}\mathbf{h}_{b,n}$, where $g_{b,n}$ is the large-scale fading component depending on the devices location, and $\mathbf{h}_{b,n} \in \mathbb{C}^{M}$ is the small-scale fading following independent and identically distributed (i.i.d.) complex Gaussian distribution with zero mean and unit variance.

The grant-free random access protocol is adopted in this paper \cite{qiang}. Specifically, at the beginning of each time slot, the active devices transmit their pilot sequences over the uplink channels simultaneously and then the APs perform the activity detection based on the received signals in a cooperative manner. All pilot sequences, $\mathbf{s}_n \in \mathbb{C}^{L}, n\in \{1,2,\cdots,N\}$, are generated from i.i.d. complex Gaussian distribution with zero mean and unit variance which are known at the APs in advance. Thus, the received signal $\mathbf{Y}_b\in \mathbb{C}^{L\times M}$ at the $b$th AP can be expressed as
\begin{eqnarray}
\label{aps}
\mathbf{Y}_b=\sum_{n=1}^N\chi_n\mathbf{s}_n \sqrt{g_{b,n}}\mathbf{h}_{b,n}^T+\mathbf{W}_b=\mathbf{S}\boldsymbol{\Gamma}_b^{\frac{1}{2}}\mathbf{H}_b+\mathbf{W}_b
\end{eqnarray}
where $\mathbf{H}_b=[\mathbf{h}_{b,1},\cdots,\mathbf{h}_{b,N}]^T\in \mathbb{C}^{N \times M}$ denotes the small-scale fading channel matrix, $\mathbf{S}=[\mathbf{s}_1,\cdots,\mathbf{s}_N]\in \mathbb{C}^{L \times N}$ denotes the horizontal stack of all pilot sequences, and $\mathbf{W}_b \in \mathbb{C}^{L \times M}$ is the additive white Gaussian noise (AWGN) marix with i.i.d. entries $~\mathcal{CN}(0,\sigma^2)$, where $\sigma^2$ denotes the noise power at each antenna. In this paper, we adopt $(\cdot)^H$ and $(\cdot)^T$ to denote conjugate transpose and transpose, respectively. Define $\boldsymbol{\gamma}_b =[\gamma_{b,1},\cdots,\gamma_{b,N}]^T\in \mathbb{R}^{N \times 1}$ as the diagonal entries of $\boldsymbol{\Gamma}_b$, representing the device state vector of the $b$th AP with $\gamma_{bn}=\chi_n g_{b,n}$. The APs detect the active devices by estimating the term $\chi_n g_{b,n}$. Especially, since the term $\boldsymbol{\Gamma}_b$ can be determined by the covariance of the received signal, we aim to design a covariance-based cooperative activity detection algorithm via a limited cooperation among multiple APs.

\section{Cooperative Massive Detection Algorithm}
In this section, we first propose a cooperative detection framework for 6G cell-free wireless networks with a massive number of IoT devices. Then, we design a corresponding cooperative detection algorithm.

\subsection{Cooperative Massive Detection Framework}
For the problem in model (\ref{aps}), the unknown device state vectors for different APs are different. Moreover, there are some common characteristics among the neighboring APs. To enhance the detection
performance, we first associate a local estimator with each AP. Incorporating the estimates of the neighboring APs, i.e., sparsity-promoting and the similarity-promoting terms \cite{shaodim, sim}, we can modify the local estimator to associate a regularized local cost function with each AP.

Firstly, we design the local estimator. It is well known that the covariance-based massive activity detection is equivalent to recovering the device state vector $\boldsymbol{\gamma}_b$ from the noisy measures $\mathbf{Y}_b$ with the knowledge of the pre-defined pilot sequence matrix $\mathbf{S}$.
In general, the estimation of the device state vector $\boldsymbol{\gamma}_b$ can be formulated as a maximum likelihood estimation problem \cite{covar}. In particular, for a given $\mathbf{Y}_b$, each column of $\mathbf{Y}_b$, denoted as $\mathbf{y}_{bm}$, $1 \leq m \leq M$, can be termed as an independent sample having the following multivariate complex Gaussian distribution:
\begin{equation}\label{ml}
\mathbf{y}_{bm}\sim \mathcal{CN}(\mathbf{0},\mathbf{S}\boldsymbol{\Gamma}_b\mathbf{S}^H+\sigma^2\mathbf{I}),
\end{equation}
where the covariance matrix is calculated by $\mathbb{E}[\mathbf{y}_{bm}\mathbf{y}_{bm}^H]$ and $\mathbf{I}$ denotes the identity matrix. For convenience, we define
$\boldsymbol{\Sigma}_b=\mathbf{S}\boldsymbol{\Gamma}_b\mathbf{S}^H+\sigma^2\mathbf{I}$.
Then, the likelihood of $\mathbf{Y}_b$ given $\boldsymbol{\gamma}_b$ can be represented as
\begin{eqnarray}\label{liki}
P(\mathbf{Y}_b|\boldsymbol{\gamma}_b)
=\frac{1}{\det(\pi\boldsymbol{\Sigma}_b)^M}\exp(-\text{tr}(\boldsymbol{\Sigma}_b^{-1}\mathbf{Y}_b\mathbf{Y}_b^H)),
\end{eqnarray}
where $\det(\cdot)$ and $\text{tr}(\cdot)$ are operators that return the determinant and the trace of a matrix, respectively.
By exploiting the Gaussianity, we can obtain the Maximum Likelihood (ML) estimator of $\boldsymbol{\gamma}_b$ at the $b$th AP as follows:
\begin{eqnarray}\label{eqr}
  f(\boldsymbol{\gamma}_b)=-P(\mathbf{Y}_b|\boldsymbol{\gamma}_b)=\ln\text{det}(\boldsymbol{\Sigma}_b)+\text{tr}(\boldsymbol{\Sigma}_b^{-1}\hat{\boldsymbol{\Sigma}}_{b\mathbf{y}}),
\end{eqnarray}
where $\hat{\boldsymbol{\Sigma}}_{b\mathbf{y}}=\frac{1}{M}\mathbf{Y}_b\mathbf{Y}_b^H$ denotes the sample covariance matrix of the received signal of the $b$th AP averaged over different antennas. Based on (\ref{eqr}), the maximum likelihood estimation problem can be formulated as $\arg \min_{\boldsymbol{\gamma}_b\in\mathbb{R}_+}f(\boldsymbol{\gamma}_b)$.

Secondly, since the activity detection is a typical sparse signal processing problem, we propose a sparsity-promoting term to facilitate cooperative detection. The specific sparsity pattern can be simultaneously observed at different APs, namely the indices of nonzero entries of $\boldsymbol{\gamma}_b$ are consistent for $b=1,2,\cdots,B$. Because each AP only communicates with its neighbor APs, it cannot obtain the global information about the sparsity pattern. Moreover, it is quite challenging to split this global quantity into several local quantities consisting of components only from the neighboring nodes. In this case, for the $b$th AP, we define a local parameter matrix consisting of the parameter vectors of all its neighbors, which can be directly obtained as follows:
\begin{equation}\label{spas1}
  \mathbf{R}_b=\left [ \boldsymbol{\gamma}_{l_1},\boldsymbol{\gamma}_{l_2},\boldsymbol{\gamma}_{l_i},\cdots,\boldsymbol{\gamma}_{l_{|\mathcal{N}_b^{-}|_c}},\boldsymbol{\gamma}_b \right ] \in \mathbb{C}^{N \times (|\mathcal{N}_b|_c)},
\end{equation}
where $l_i\in \mathcal{N}_b^{-}$ is the index set of neighbors of the $b$th AP except itself, $\mathcal{N}_b$ denotes the index set of the neighbors of the $b$ AP including itself, and $|\mathcal{N}_b^{-}|_c$ and $|\mathcal{N}_b|_c$ denote the cardinality of the set $\mathcal{N}_b^{-}$ and $\mathcal{N}_b$, respectively. Consequently, we aim to impose sparsity constraints on the row vectors of matrix $\mathbf{R}_b$ to exploit the joint sparsity. To this end, this paper designs a novel
sparsity-promoting term, which is given by
\begin{equation}\label{spas}
   g(\boldsymbol{\gamma}_b)=\sum_{n=1}^{N}\left(\left \| \mathbf{R}_b(n,:) \right \|_2-\frac{1}{\theta}\ln(1+\theta\left \| \mathbf{R}_b(n,:) \right \|_2)\right),
\end{equation}
where $\theta>0$ is the penalty parameter, $\mathbf{R}_b(n,:)$ is the $n$th row of $\mathbf{R}_b$, and $\left \| \cdot \right \|_{2}$ denotes the $l_2$ norm of a matrix. Herein, $g(\boldsymbol{\gamma}_b)$ is the logarithmic smooth function which can promote row sparsity \cite{shaodim}, where the nonzero rows are penalized by minimizing $g(\boldsymbol{\gamma}_b)$. In this way, a common sparsity profile across the columns of the local parameter matrix $\mathbf{R}_b$ is promoted. Although the sparsity-promoting term is imposed on the local parameter matrix $\mathbf{R}_b$, the cooperative nature promotes a common sparsity profile across all columns of the global device state vectors $\{\boldsymbol{\gamma}_b\}_{b=1}^B$.

Thirdly, we design a similarity-promoting term to improve the detection performance. The supports of the global device state vector $\{\boldsymbol{\gamma}_b\}_{b=1}^B$ for all APs are the same, but the amplitudes of the nonzero entries at the APs are different from each other due to the effects of different path loss. In particular, the device state vectors of neighboring APs have a large number of similar entries and only a relatively small number of distinct entries. Motivated by these observations, we design a similarity-promoting function as follows
\begin{equation}\label{sim}
\Psi(\boldsymbol{\gamma}_b)=\sum_{l\in \mathcal{N}_b}c_{lb}\Psi_l(\boldsymbol{\gamma}_b-\boldsymbol{\gamma}_l),
  \end{equation}
where $c_{lb}$ are linear weights satisfying the conditions:
$
\sum\limits_{l \in {\mathcal{N}_b}} {c_{lb}} = 1, ~~{c_{lb}} = 0~~ \forall l \notin {\mathcal{N}_b}$.
$\Psi_l(\boldsymbol{\gamma}_b-\boldsymbol{\gamma}_l)$ is a convex penalty function, minimized at $\Psi_l(\mathbf{0})$, which encourages similarity between $\boldsymbol{\gamma}_b$ and $\boldsymbol{\gamma}_l$. Note that the log-likelihood $f(\boldsymbol{\gamma}_b)$ depends on the empirical covariance $\hat{\boldsymbol{\Sigma}}_{b\mathbf{y}}$. In high-dimensional settings, where
the length of pilot sequences $L$ is larger than the number of AP antennas $M$, $\hat{\boldsymbol{\Sigma}}_{b\mathbf{y}}$ will be relatively different from the covariance matrix $\boldsymbol{\Sigma}_b$. By enforcing structural similarity, each $\boldsymbol{\Sigma}_b$ can exploit from the fact that neighboring AP estimates should be similar to each other.

The specific expressions in the penalty function $\Psi_l(\cdot)$ form can be set different. In general, it dependents on the assumptions imposed on the problem, one may choose the most appropriate penalty for the data at hand. For example, $l_1$-norm penalty $\Psi_l(\mathbf{x})=\sum_{n=1}^N|x_n|$, where $x_n$ and $|\cdot|$ denote the $n$th element of the vector $\mathbf{x}$ and the absolute value, respectively. This penalty function encourages the changes of limited number of values between neighbor APs, while the rest of the structure remains the same. In other words, it borrows information aggressively across neighbors, encouraging not only similar structure but also similar values. As a result, this
penalty is suitable for massive access where only a small fraction of potential devices to change their states at a time slot and is adopted in this paper.

After defining the similarity-promoting term and sparsity-promoting term, accumulating them into (\ref{eqr}) leads to the following novel regularized local cost function at the $b$th AP:
\begin{eqnarray}\label{ob1}
F(\boldsymbol{\gamma}_b)\!=\!f(\boldsymbol{\gamma}_b)
+ \beta g(\boldsymbol{\gamma}_b)+\tau\Psi(\boldsymbol{\gamma}_b), ~\forall b\!\in\! \{1,2,\cdots,B\},
\end{eqnarray}
where $\beta>0$ and $\tau>0$ are the penalty parameters used to enforce sparsity and similarity, respectively. In the following, we design a massive activity detection algorithm to minimize the local cost function at each AP.

\subsection{A Decentralized Approximate Separating Strategy}
Note that the first term of \eqref{ob1}, i.e., $f(\boldsymbol{\gamma}_b)$ is differentiable and geodesically convex \cite{geo}. However, as stated in the above subsection, $\Psi_l(\boldsymbol{\gamma}_b-\boldsymbol{\gamma}_l)$ is discontinuous, i.e., the third term of the local cost function could be a sum of non-smooth functions, and the second term is also potentially non-differentiable. In addition, the unknown variables $\boldsymbol{\gamma}_b$ for neighboring APs are coupled with each other. These obstacles make the problem intractable to solve and existing algorithms are not applicable to such a problem. In the following, we design a decentralized approximate separating strategy for minimizing the cost function in \eqref{ob1} based on the forward-backward splitting strategy \cite{forward}, which can handle the non-smooth problem and is especially amenable to solve the high-dimensional activity detection problem due to its fast convergence rate and its conceptual and mathematical simplicity.

Before proceeding, we recall the forward-backward splitting approach for minimizing \eqref{ob1}, which is given by the iteration
\begin{equation}\label{gbp}
  \boldsymbol{\gamma}_b^{t+1}=\text{prox}_{\eta_b(\tau\Psi+\beta g)}(\boldsymbol{\gamma}_b^{t}-\eta_b\bigtriangledown f(\boldsymbol{\gamma}_b^t)),
\end{equation}
where $\bigtriangledown f(\cdot)$ is the gradient of function $f$, $\eta_b$ is the step size for the $b$th AP, and $\boldsymbol{\gamma}_b^t$ denotes the value of $\boldsymbol{\gamma}_b$ in the $t$th iteration. The gradient descent
step is the forward step and the proximal step is the backward step \cite{forward}.
Note that the proximal operator of a function $h$ is a mapping function given by:
$\text{prox}_{\eta h}(\mathbf{y})=\arg\min_{\mathbf{u}} h(\mathbf{u})+\frac{1}{2\eta}\left \| \mathbf{u}-\mathbf{y} \right \|_2^2$
with variables $\mathbf{y}$ and $\mathbf{u}$, and a step-size $\eta > 0$ \cite{acce}.

Unfortunately, it is prohibitively challenging to directly evaluate the proximal operators with respect to similarity-promoting function $\Psi(\boldsymbol{\gamma}_b)$ and the sum of $\beta g(\boldsymbol{\gamma}_b) + \tau\Psi(\boldsymbol{\gamma}_b)$. Moreover, the calculation of $\Psi(\boldsymbol{\gamma}_b)$ over all the number of neighborhood, $|\mathcal{N}_b|_c$, in each iteration is expensive. Motivated by Douglas Rachford splitting in \cite{forward}, where two of the proximal operators can be updated alternately, this paper aims to handle the proximal operator of function $\Psi(\boldsymbol{\gamma}_b)$ and $g(\boldsymbol{\gamma}_b)$ separately. Specifically, we first calculate an estimator $\mathbf{x}_b^t$ of the subgradient $\partial\Psi(\boldsymbol{\gamma}_b^t)$ and then incorporate the gradient descent step into the proximal step with respect to sparsity-promoting term $g(\cdot)$ for the $b$th AP, which is given by
\begin{equation}\label{zt}
  \mathbf{z}_b^{t}=\text{prox}_{\beta\eta_b g}(\boldsymbol{\gamma}_b^t-\eta_b\bigtriangledown f(\boldsymbol{\gamma}_b^t)-\tau\eta_b\mathbf{x}_b^t),
\end{equation}
where $\mathbf{z}_b^{t}$ is a intermediate variable. Then, according to the update rule of Douglas Rachford splitting, we incorporate $\mathbf{z}_b^{t}$ into proximal operator with respect to similarity-promoting function $\Psi(\boldsymbol{\gamma}_b)$:
\begin{eqnarray}\label{DR}
  \boldsymbol{\gamma}_b^{t+1}=\text{prox}_{\tau\eta_b \Psi}(\mathbf{z}_b^t+\tau\eta_b \mathbf{x}_b^{t}).
\end{eqnarray}

The intermediate variable $\mathbf{z}_b^{t}$ and device state vector $\boldsymbol{\gamma}_b^{t}$ iterate alternately and their values approach to each other. When converging to optimality, their values are identical. Afterwards, in order to overcome the difficulty in processing the non-smooth finite sum term and to reduce the computational overhead, this paper proposes a splitting strategy that uses the proximal operator of a single function $\Psi_l$ in each iteration to approximate the proximal operator of the average of $|\mathcal{N}_b|_c$ non-smooth functions $\Psi_l$. In mathematical terms, we first choose $l$ randomly from the set $\mathcal{N}_b$ with probabilities $\{p_1,p_2,\cdots,p_{\left | \mathcal{N}_b \right |_c}\}$. Then, utilizing the proximal operator, a specific step is introduced as follows
\begin{equation}\label{gb}
  \boldsymbol{\gamma}_b^{t+1}=\text{prox}_{\tau\eta_b^l \Psi_l}(\mathbf{z}_b^t+\tau\eta_b^l \mathbf{x}_b^{l,t}),
\end{equation}
where $\mathbf{x}_b^{l,t}$ is the estimator of subgradient $\partial\Psi_l(\boldsymbol{\gamma}_b^{t+1})$ for the randomly selected $l$th neighbor of the $b$th AP in the $t$th iteration. Let
$c_{lb}^t$ denotes the combiner at $t$th iteration. In the sequel, $\eta_b^{l}$ can be set to $\eta_b^{l}=\frac{c_{lb}^t\eta_b}{ p_l}$, which is a stochastic approximation of $\eta_b$ controlled by the combiner and the probability of being selected. In this way, we are able to treat the difficult term in (\ref{ob1}) with non-smooth finite sum term for any size of cardinality $|\mathcal{N}_b|_c$.

Since $\mathbf{z}_b^{t}$ and $\boldsymbol{\gamma}_b^{t}$ converge to the same value,
\eqref{gb} is an accurate approximation of \eqref{gbp} if $\mathbf{x}_b^{l,t}=\partial\Psi_l(\boldsymbol{\gamma}_b^{t+1})$ and $\mathbf{x}_b^{t}=\partial\Psi(\boldsymbol{\gamma}_b^{t+1})$ hold. Thus, we must ensure that $\partial\Psi_l(\boldsymbol{\gamma}_b^{t+1})$ is close to $\mathbf{x}_b^{l,t}$ to obtain an accurate estimator. According to the definition of proximal operator, equation \eqref{gb} satisfies
\begin{eqnarray}\label{xwl}
\!\!\!\!\!\!\!\!\!\!\!\!\!\!&&\!\!\!\!\!\!\!\!\!\!\!\!\frac{(\mathbf{z}_b^t\!+\!\tau\eta_b^{l} \mathbf{x}_b^{l,t}\!-\!\text{prox}_{\tau\eta_b^{l} \Psi_l}(\mathbf{z}_b^t\!+\!\tau\eta_b^{l} \mathbf{x}_b^{l,t}))}{\tau\eta_b^{l}}\in \partial\Psi_l(\boldsymbol{\gamma}_b^{t+1}).
\end{eqnarray}
Hence, we can arrive at the following subgradient estimator $\mathbf{x}_b^{l,t+1}$ such that \eqref{gb} holds:
\begin{equation}\label{xbl}
  \mathbf{x}_b^{l,t+1}=\mathbf{x}_b^{l,t}+\frac{1}{\tau\eta_b^{l}}(\mathbf{z}_b^t-\boldsymbol{\gamma}_b^{t+1}),
\end{equation}
where the right hand side (RHS) of \eqref{xbl} is obtained by replacing the proximal step in \eqref{xwl} by $\boldsymbol{\gamma}_b^{t+1}$ and further reorganizing the left hand side (LHS) of formula \eqref{xwl}. Consequently, the subgradient estimator $\mathbf{x}_b^t$ in \eqref{zt} can be updated as
\begin{equation}\label{full}
  \mathbf{x}_b^{t+1}=\mathbf{x}_b^{t}+c_{lb}^t(\mathbf{x}_b^{l,t+1}-\mathbf{x}_b^{l,t}),
\end{equation}
which exploits the fact that $ \mathbf{x}_b^{t}=\sum_{l=1}^{|\mathcal{N}_b|_c}c_{lb}^t\mathbf{x}_b^{l,t}$ and as stated in \eqref{gb}, only a single selected $\mathbf{x}_b^{l,t}$ is updated in each iteration.

\subsection{Derivation of Proximal Operators and Combiners}
Since the proximal operator needs to be calculated at each iteration in (\ref{zt}) and (\ref{gb}), it is important to derive closed form expressions for evaluating them exactly.
According to the well-known Sherman-Morrison rank-1 update identity \cite{sher}, we obtain
\begin{eqnarray}\label{sdd}
\!\!\!\!\!\!&&\!\!\!\!\!\!\left ( \boldsymbol{\Sigma}_b- {\gamma}_{bn}\mathbf{s}_n\mathbf{s}_n^H+{\gamma}_{bn}\mathbf{s}_n\mathbf{s}_n^H\right )^{-1}=\boldsymbol{\Sigma}_{bn}^{-1}-
\frac{{\gamma}_{bn}\boldsymbol{\Sigma}_{bn}^{-1}\mathbf{s}_n\mathbf{s}_n^H\boldsymbol{\Sigma}_{bn}^{-1}}{1+{\gamma}_{bn}\mathbf{s}_n^H\boldsymbol{\Sigma}_{bn}^{-1}\mathbf{s}_n},
\nonumber\\
\!\!\!\!\!\!&&\!\!\!\!\!\!
\end{eqnarray}
with $\boldsymbol{\Sigma}_{bn}=\boldsymbol{\Sigma}_b- {\gamma}_{bn}\mathbf{s}_n\mathbf{s}_n^H$, where ${\gamma}_{bn}$ is the $n$th element of $\boldsymbol{\gamma}_{b}$. Applying the well-known determinant identity yields
\begin{eqnarray}\label{sdd1}
\text{det}(\boldsymbol{\Sigma}_{bn}+{\gamma}_{bn}\mathbf{s}_n\mathbf{s}_n^H)=(1+{\gamma}_{bn}\mathbf{s}_n^H\boldsymbol{\Sigma}_{bn}^{-1}\mathbf{s}_n)\text{det}(\boldsymbol{\Sigma}_{bn}).
\end{eqnarray}
Then, substituting (\ref{sdd}) and (\ref{sdd1}) into (\ref{eqr}) and taking the derivative of $f(\boldsymbol{\gamma}_b)$ with respect to ${\gamma}_{bn}$, we have
\begin{eqnarray}\label{gsi}
\bigtriangledown f(\gamma_{bn})=\frac{\mathbf{s}_n^H\boldsymbol{\Sigma}_{bn}^{-1}\mathbf{s}_n}{1+{\gamma}_{bn}\mathbf{s}_n^H\boldsymbol{\Sigma}_{bn}^{-1}\mathbf{s}_n}-\frac{\mathbf{s}_n^H\boldsymbol{\Sigma}_{bn}^{-1}\hat{\boldsymbol{\Sigma}}_{b\mathbf{y}}\boldsymbol{\Sigma}_{bn}^{-1}\mathbf{s}_n}{(1+{\gamma}_{bn}\mathbf{s}_n^H\boldsymbol{\Sigma}_{bn}^{-1}\mathbf{s}_n)^2}.
\end{eqnarray}

Correspondingly, the gradient $\bigtriangledown f(\boldsymbol{\gamma}_b^t)$ can be derived by computing the following derivative:
$
\bigtriangledown f(\boldsymbol{\gamma}_{b}^t)=\text{col}\{\bigtriangledown f(\gamma_{b1}^t), \cdots,\bigtriangledown f(\gamma_{bN}^t)\}$,
where $\text{col}(\cdot)$ denotes a column vector.
By substituting it into (\ref{zt}) and calculating the  closed
form expression of the proximal operator of $g(\boldsymbol{\gamma}_b)$, we can obtain the following intermediate recursion
\begin{eqnarray}\label{zbt}
\!\!\!\!\!\!\!\!\!\!\!\!&&\!\!\!\!\!\!\!\!\!\!\!\! \mathbf{z}_b^{t}= \boldsymbol{\varsigma}_{b}^t-\eta_b\beta \text{col}\left\{ \frac{ \varsigma_{b1}^t}{ \left \| \mathbf{R}_b^t(1,:) \right \|_2 },\cdots,\frac{ \varsigma_{bN}^t}{ \left \| \mathbf{R}_b^t(N,:) \right \|_2 }\right\},
\end{eqnarray}
with $
  \boldsymbol{\varsigma_{b}}^t= \boldsymbol{\gamma}_b^t-\eta_b\bigtriangledown f(\boldsymbol{\gamma}_b^t)-\tau\eta_b\mathbf{x}_b^t$,
where $\varsigma_{bn}^t$ is the $n$th element of $\boldsymbol{\varsigma}_{b}^t$.

Now, we turn to derive the recursion of $\boldsymbol{\gamma}_b^{t}$ in \eqref{gb}. Since $\Psi_l(\cdot)$ in \eqref{gb} is fully separable, its proximal operator can be evaluated component-wise:
\begin{eqnarray}\label{pos1}
\gamma_{bn}^{t+1}= \left\{\begin{array}{l}
-\min\left(\tau \eta_b^l\frac{z_{bn}^t+\tau\eta_b^l x_{bn}^{l,t}-\boldsymbol{\gamma}_{ln}^t}{|z_{bn}^t+\tau\eta_b^l x_{bn}^{l,t}-\boldsymbol{\gamma}_{ln}^t|},z_{bn}^t+\tau\eta_b^l x_{bn}^{l,t}\right)\\
+z_{bn}^t+\tau\eta_b^l x_{bn}^{l,t},~~~~\text{if}~~ z_{bn}^t+\tau\eta_b^l x_{bn}^{l,t}\neq0,\\
0,~~~~~~~~~~~~~~~~~~~~\text{if}~~ z_{bn}^t+\tau\eta_b^l x_{bn}^{l,t}=0,
\end{array} \right.
\end{eqnarray}
where the minimization operator is to preserve the positivity of $\gamma_{bn}$. Here, $z_{bn}^t$ and $x_{bn}^{l,t}$ are the $n$th entry of vectors $\mathbf{z}_b^t$ and $\mathbf{x}_b^{l,t}$, respectively. For (\ref{full}) and (\ref{pos1}), the estimation performance depends, to a great extent, on the cooperation strategy specified by the combiner $c_{lb}^t$. This paper adopts the following adaptive combiner
\begin{eqnarray}\label{clb}
c_{lb}^t = \left\{\begin{array}{l}
\frac{2}{\left | \mathcal{N}_b^{-} \right |_c}\frac{1}{1+\exp(\rho \|\boldsymbol{\gamma}_b^{t-1}-\boldsymbol{\gamma}_l^{t-1}\|_2 )},~l\in \mathcal{N}_b^{-},\\
1-\sum _{l\in \mathcal{N}_b^{-}}c_{lb}^t,~~~~~~~~~~~~~~~~l=b,\\
0, ~~~~~~~~~~~~~~~~~~~~~~~~~~~~~~~~l\notin  \mathcal{N}_b,
\end{array} \right.
\end{eqnarray}
where $\rho$ is a large constant set beforehand. Note that the term $\|\boldsymbol{\gamma}_b^{t-1}-\boldsymbol{\gamma}_l^{t-1} \|_2$ in (\ref{clb}) accounts for the distance between the local estimates of the $b$th AP and its $l$th neighbor. The combiner $c_{lb}^t$ is inversely proportional to such a distance. When the distance defined above between two APs is large, the $b$th AP tends to decrease the combination weight, or even discard the information from this neighbor. Conversely, the $b$th AP will increase the combination weight when the distance of estimation between two APs is small. For clarity, the pseudo-code of the CMD algorithm is summarized in Algorithm 1.

Once an estimate $\{\boldsymbol{\gamma}_b\}_{b=1}^B$ is obtained, we employ the element-wise thresholding at each AP to determine $\chi_n$ from $\gamma_{bn}$ which is the $n$-th entry of $\boldsymbol{\gamma}_b$. In specific, $\chi_n=1$ if $\gamma_{b^0 n}>\imath\sigma^2$ for a pre-specified threshold $\imath>0$, and $\chi_n=0$ otherwise \cite{mmwave}. Herein, $b^0$ is the AP closest to device $n$. Note that there exists a particular fixed point $\boldsymbol{\gamma}_b^{*}$ of the problem $\arg \min_{\boldsymbol{\gamma}_b\in\mathbb{R}_+}F(\boldsymbol{\gamma}_b)$
for \eqref{zt} and \eqref{gb}. Based on the information across the 6G wireless network, it can be shown that the iterates $\boldsymbol{\gamma}_b^{t}$ converge to this particular fixed point under certain step-size conditions.

\begin{algorithm}[h]
\caption{Cooperative Massive Detection Algorithm}
\label{alg1}
\begin{algorithmic}[1]
\STATE \textbf{Input}: $\{\mathbf{Y}_b\}_{b=1}^B$, $\mathbf{S}$, $\{\hat{\boldsymbol{\Sigma}}_{b\mathbf{y}}=\frac{1}{M}\mathbf{Y}_b\mathbf{Y}_b^H\}_{b=1}^B$, step size $\{\eta_b\}_{b=1}^B$, and total iterations $T$.
\STATE \textbf{Initialization}: $\{\boldsymbol{\gamma}_b^{0}=\mathbf{0}\}_{b=1}^B$, $\{\boldsymbol{\Sigma}_b^0=\sigma^2\mathbf{I}\}_{b=1}^B$, $\{\mathbf{x}_b^{l,0}, l\in \mathcal{N}_b\}_{b=1}^B$, $\{\mathbf{x}_b^0= \sum_{l\in \mathcal{N}_b}c_{lb}^0\mathbf{x}_b^{l,0}\}_{b=1}^B$. \\
\FOR{$t=1 : T$}
\STATE {\bf{for each AP $b$:}}
\STATE {\bf{Adaptation:}}
\STATE Compute $\mathbf{z}_b^{t}
$ based on (\ref{zbt})
\STATE Choose $l$ randomly from the set $\mathcal{N}_b$ with probabilities $\{p_1,p_2,\cdots,p_{\left | \mathcal{N}_b \right |_c}\}$
\STATE Compute adaptive combiner $c_{lb}^t$ based on (\ref{clb})
\STATE Compute $\eta_b^{l}=\frac{c_{lb}^t\eta_b}{ p_l}$
\FOR{$n=1 : N$}
\STATE  Update $\gamma_{bn}^{t+1}$ based on (\ref{pos1})
\STATE $\boldsymbol{\Sigma}_b^{t+1}=\boldsymbol{\Sigma}_b^{t}+(\gamma_{bn}^{t+1}-\gamma_{bn}^{t})\mathbf{s}_n\mathbf{s}_n^H$
\ENDFOR
\STATE Compute $\mathbf{x}_b^{l,t+1}$ based on \eqref{xbl}
\STATE Compute $\mathbf{x}_b^{t+1}$ based on \eqref{full}
\STATE {\bf{Communication:}}
\STATE Transmit $\boldsymbol{\gamma}_{b}^t$ to its one-hop neighbor AP
\ENDFOR
\STATE \textbf{Output}: $\{\boldsymbol{\gamma}_b^{t+1}\}_{b=1}^B$
\end{algorithmic}
\end{algorithm}

\subsection{Computational Complexity and Communication Cost}
In what follows, the computational complexity and communication
cost of the proposed CMD algorithm is analyzed. In each iteration, for an arbitrary AP, the computational complexity mainly arises from the matrix multiplication, and the overall computational complexity of the CMD algorithm is $\mathcal{O}(L^2N)$.
Although the computational complexity of sample covariance $\hat{\boldsymbol{\Sigma}}_{b\mathbf{y}}$ is $\mathcal{O}(L^2M)$, it only needs to be calculated once at each time slot before the iteration. For the communication cost, in each iteration, each
AP needs to transmit $N$-dimensional intermediate $\boldsymbol{\gamma}_{b}^t$ to its neighboring APs. Thus, for all APs, the CMD algorithm needs to exchange $N\sum_{b=1}^B\left|\mathcal{N}_b^{-}\right|_c$ parameters.
Since the APs exchange intermediate variables instead of the received signal matrix $\mathbf{Y}_b$, the communication cost does not grow as the number $M$ of AP antennas increases.

\emph{Remark 1}: It is interesting to emphasize that the computational complexity and the communication cost at each iteration of the CMD algorithm do not grow as the number of each AP antennas, $M$, increases. Note that the CMD algorithm can also be adopted for data detection in unsourced random access \cite{unsourced}, where an arbitrary AP only needs to know which messages are sent without identifying which message belongs to which device. Suppose that each active device has $q$ bits to send, then the total number of potential devices $N$ in the device activity detection problem is replaced by $2^J$, where the small size $J$ is obtained by divided $q$-bit message into $\mathcal{Z}$ blocks. The details please refer to our journal paper. Thus, the computational cost of CMD algorithm reduces to $2^J\sum_{b=1}^B\left|\mathcal{N}_b^{-}\right|_c$. This implies that the communication cost of the proposed CMD algorithm does not grow by increasing the total number of potential devices, which is an appealing feature for reality IoT networks.

\section{Numerical Results}
In this section, we present numerical simulations to validate the effectiveness of the proposed CMD algorithm. We simulate the underdetermined 6G cell-free wireless network comprising $B=20$ APs geographically distributed in a vast area to serve $N$ potential devices. The AP-to-AP distance is $500$ m and the number of connections between APs can be set differently. The positive constants $\theta$ is selected to be $1/0.039$. The
penalty parameters $\beta$ and $\tau$ are set as $0.038$ and $0.0075$,
respectively. The step size $\eta_b=0.003$ and is the same for all APs, $p_l= \frac{1}{|\mathcal{N}_b|_c}$, and $\rho=500$.

As a performance measure, we adopt the activity error rate (AER). The AER is a sum of the missed detection probability, defined as the probability that a device is active but is declared to be inactive, and the false-alarm probability, defined as the probability that a device is inactive but the detector declares it to be active. As a reference, we compare the proposed CMD algorithm with two baseline schemes: the conventional ML-based multi-cell algorithm \cite{covar} and the AMP-based multi-cell algorithm, where each AP only serves its cell's devices without multi-cell cooperation and treats the inter-cell interference as noise \cite{multicell}.

\begin{figure}[h]
  \centering
\includegraphics [width=0.29\textwidth] {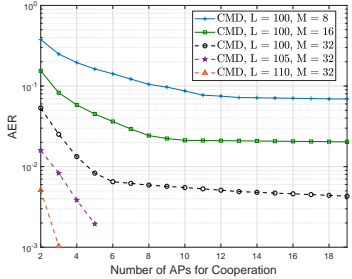}
\caption{The AER for different number of APs for cooperation with potential devices $N = 1,000$, active device $K=200$, and SNR is set as 10 dB.}
\label{edgedegree}
\end{figure}

Fig. \ref{edgedegree} depicts the detection performance versus different choices of the number of APs for cooperation. Initially, in the area with a few numbers of cooperation APs, the AER decreases sharply as the number of cooperation APs increases. When the number of cooperation APs continues to increase, the performance improvement diminishes. In addition, it is observed that such a performance saturation point value depends on AP antennas $M$ and pilot length $L$, i.e., increasing $M$ or $L$ helps decrease the saturation point value, which indicates that the AP becomes more capable of
detecting the activity with a low communication cost. The reason for this phenomenon is that for an arbitrary AP, more connections result in more intermediate estimates exchange in the proposed CMD algorithm, leading to good detection performance. However, the channel strengths from a specific active device to the far away APs are approximate zero, and the intermediate estimates exchange with the remote AP can not further improve the massive detection performance significantly. The results also indicate that only a small number of APs are required for cooperation which strikes a tradeoff between the detection performance and the communication cost.

\begin{figure}[htbp]
\begin{minipage}[t]{0.52\linewidth}
\centering
\includegraphics[width=1.87in]{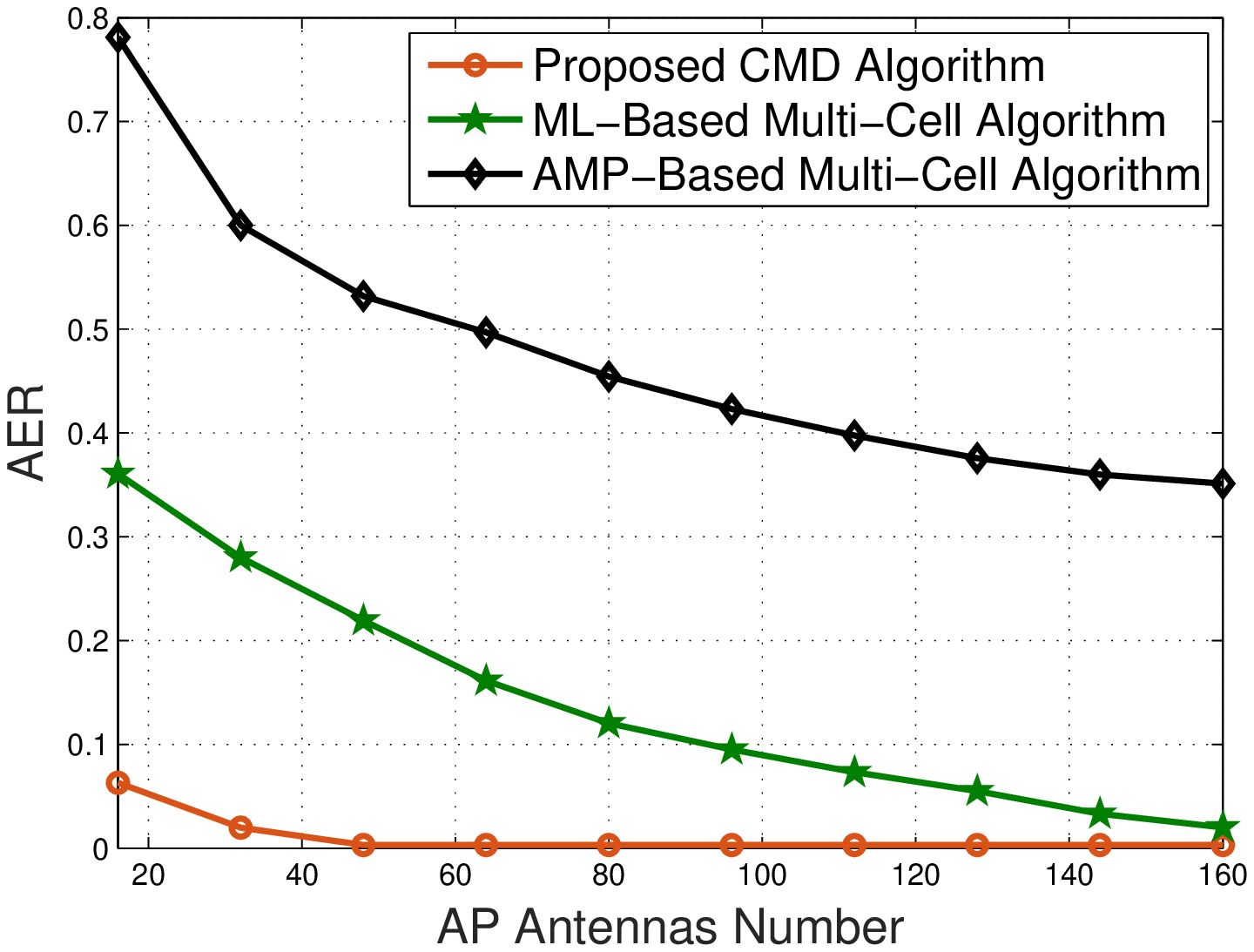}
\caption{The AER versus $M$.}
\label{antenna}
\end{minipage}%
\begin{minipage}[t]{0.52\linewidth}
\centering
\includegraphics[width=1.87in]{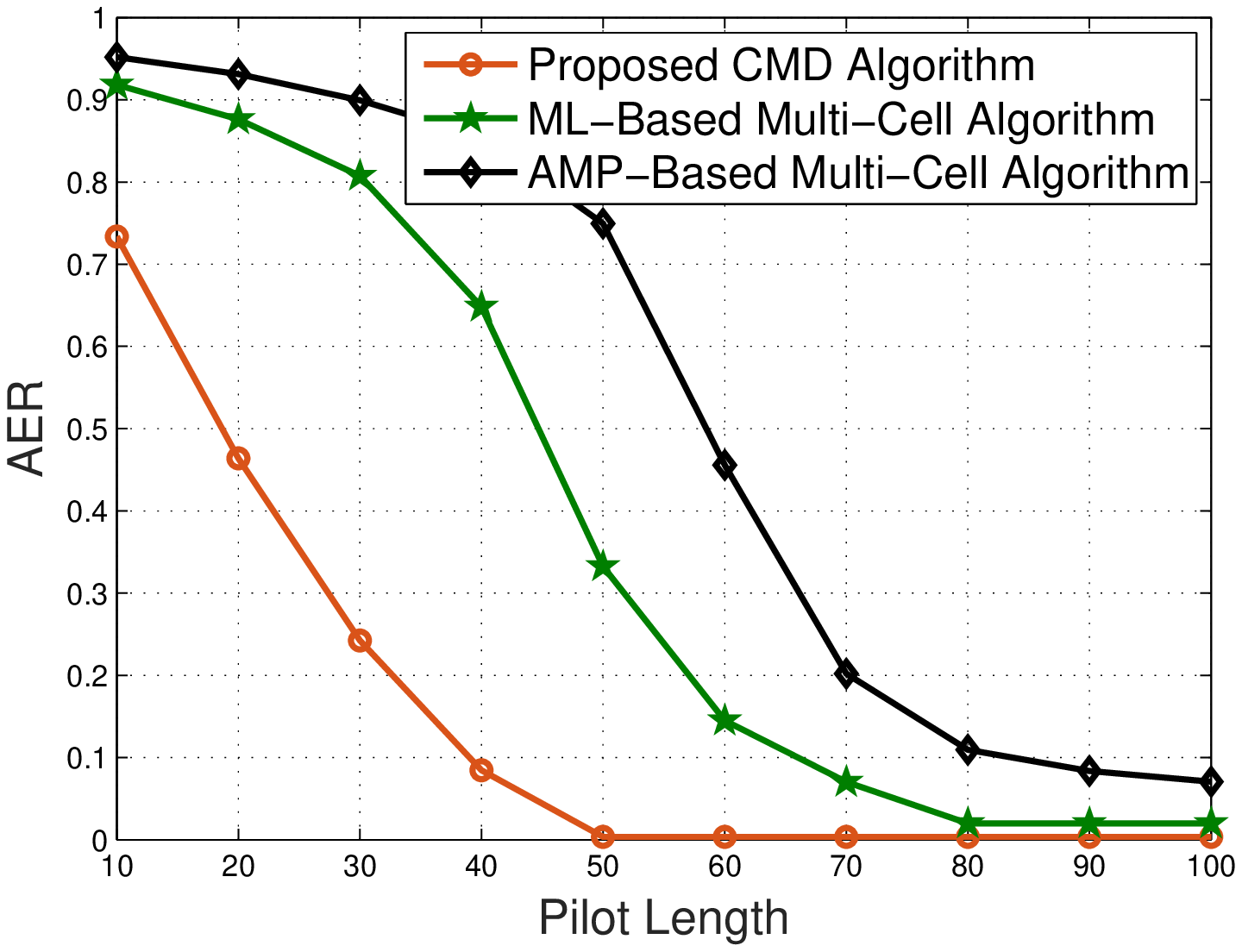}
\caption{The AER versus $L$.}
\label{pilot}
\end{minipage}%
\end{figure}

In the rest of the simulations, the number of APs for cooperation is
set to $5$. Fig. \ref{antenna} demonstrates the activity detection performance versus different numbers of AP antennas $M$ with potential devices $N = 1,000$, active devices $K=200$, pilot length $L=100$, and SNR is set as 10 dB. It is seen that the proposed CMD algorithm provides much lower AER than that of the baseline ones and the performance gap is enlarged as the number of AP antennas
increases. In other words, the superiority of the proposed CMD algorithm is evident in massive MIMO systems, which is a key technique for 6G wireless networks. Such an advantage of cooperative strategies mainly comes from that the proposed algorithm exploits the joint sparsity and similarities of the multiple APs, and the closed-form expressions for the proximal operators are derived to achieve higher efficiency. In contrast, ML-based and AMP-based multi-cell approaches ignore such prior information and only perform activity detection for the devices distributed in its own cell, where the inter-cell interference is also a severely limiting factor for reliable activity detection.

Fig. \ref{pilot} shows the detection performance versus the length of the pilot sequence $L$ with potential devices $N = 500$, active devices $K=100$, $M = 32$ antennas at the AP, and SNR is set as 10 dB. From this figure, we observe that the activity detection performance of all the considered algorithms increases as the pilot length increases and the CMD algorithm achieves a substantial performance gain over the ML-based multi-cell algorithm and the AMP-based multi-cell algorithm. Note that the CMD algorithm does not require the knowledge of channel strengths which only needs to estimate a smaller number of unknown parameters, thus, it is more efficient for activity detection than that of the AMP-based multi-cell detection approach. We can also see that the performance gap between the proposed algorithm and the baseline ones is large, especially when the pilot sequence is short.

\section{Conclusion}
This paper designed a grant-free cooperative random access framework for mMTC in 6G cell-free wireless networks based on the covariance of the received signals. By exploiting the special characteristic of the device state vectors of interest, we developed a covariance-based high-accuracy and low-complexity algorithm. Simulation results showed that the proposed algorithm can almost achieve near-optimal activity detection performance.

\end{document}